# Exceptional fracture toughness of CrCoNi-based medium- and high-entropy alloys close to liquid helium temperatures

Dong Liu[a], Qin Yu[b], Saurabh Kabra[c], Ming Jiang[a], Paul Forna-Kreutzer[a], Ruopeng Zhang[d,e], Madelyn Payne[d,e], Flynn Walsh[b,d], Bernd Gludovatz[f], Mark Asta[b,d], Andrew M. Minor[d,e], Easo P. George[g,h], Robert O. Ritchie[b,d,*]

[a]School of Physics, University of Bristol, Bristol, BS8 1TL, UK
[b]Materials Sciences Division, Lawrence Berkeley National Laboratory, Berkeley, California 94720, USA
[c]ENGIN-X, ISIS Facility, Rutherford Appleton Laboratory, Harwell campus, Oxon OX110QX, UK
[d]Department of Materials Science & Engineering, University of California, Berkeley, California, 94720, USA
[e]National Center for Electron Microscopy, Molecular Foundry, Lawrence Berkeley National Laboratory, Berkeley, CA 94720, USA
[f]School of Mechanical and Manufacturing Engineering, University of New South Wales (UNSW Sydney), Sydney, NSW 2052, Australia
[g]Materials Science & Technology Division, Oak Ridge National Laboratory, Oak Ridge, Tennessee 37831, USA
[h]Materials Science & Engineering Department, University of Tennessee, Knoxville, Tennessee 37996, USA

* Corresponding author (roritchie@lbl.gov)

**Medium- and high-entropy alloys based on the CrCoNi-system have been shown to display outstanding strength, tensile ductility and fracture toughness (damage-tolerance properties), especially at cryogenic temperatures. Here we examine the $J_{Ic}$ and (back-calculated) $K_{JIc}$ fracture toughness values of the face-centered cubic, equiatomic CrCoNi and CrMnFeCoNi alloys at 20 K. At flow stress values of ~1.5 GPa, crack-initiation $K_{JIc}$ toughnesses were found to be exceptionally high, respectively 235 and 415 MPa√m for CrMnFeCoNi and CrCoNi, with the latter displaying a crack-growth toughness $K_{ss}$ exceeding 540 MPa√m after 2.25 mm of stable cracking, which to our knowledge is the highest such value ever reported. Characterization of the crack-tip regions in CrCoNi by scanning electron and transmission electron microscopy reveal deformation structures at 20 K that are quite distinct from those at higher temperatures and involve heterogeneous nucleation, but restricted growth, of stacking faults and fine nano-twins, together with transformation to the hexagonal closed-packed phase. The coherent interfaces of these features can promote both the arrest and transmission of dislocations to generate respectively strength and ductility which strongly contributes to sustained strain hardening. Indeed, we believe that these nominally single-phase, concentrated solid-solution alloys develop their fracture resistance through a progressive synergy of deformation mechanisms, including dislocation glide, stacking-fault formation, nano-twinning and eventually *in situ* phase transformation, all of which serve to extend continuous strain hardening which simultaneously elevates strength and ductility (by delaying plastic instability), leading to truly exceptional resistance to fracture.**

**One Sentence Summary:** CrCoNi-based medium- and high-entropy alloys display exceptionally high fracture toughness near liquid-helium temperatures due to an extended sequence of deformation mechanisms that enhance both strength and ductility to prolong strain hardening.

High-entropy alloys (HEAs) have attracted increasing attention in the metallurgy community as a new class of metallic materials that derive their properties from the presence of multiple principal elements, rather than from a single dominant constituent as in most traditional metallic alloys (*e.g.*, Fe in steels). Inspired by two seminal papers,[1,2] the field has grown to encompass equiatomic as well as non-equiatomic alloys, single-phase solid solutions as well as multiphase compositionally complex alloys, with the goal of finding unique combinations of mechanical and functional properties compared to conventional alloys.[3–7]

One prominent group of such materials are the single-phase, face-centered cubic (*fcc*), equiatomic alloys based on the Cr-Co-Ni system. Among these, the equiatomic CrMnFeCoNi (so-called Cantor) alloy is the most characterized of all HEAs.[1,6-11] It came to prominence because its room-temperature mechanical properties of strength, ductility, and toughness can be significantly enhanced at liquid-nitrogen temperatures.[8-10] Specifically, its tensile strength was found to increase, in a 6 μm grain size alloy, from 763 MPa at 293 K to 1.28 GPa at 77 K, with a corresponding increase in tensile ductility from 57% to over 71%; more notably, its crack-initiation fracture toughness, $K_{JIc}$, remained at roughly 220 MPa√m over this entire temperature range, with a crack-growth toughness, $K_{ss}$, of over 300 MPa√m (after 2.25 mm of crack extension).[10] More recent experiments on this HEA showed a similar increase in strength and toughness (the latter measured in terms of the absorbed deviatoric strain energy) when the strain rate was increased from $10^{-3}$ s$^{-1}$ (quasi-static compression) up to extremely high rates of 6 x $10^5$ s$^{-1}$ (dynamic shear).[11]

There have been several derivatives of the Cantor alloy,[12,13] most notably the single-phase equiatomic CrCoNi medium-entropy alloy (MEA), which displays even better properties. At 77 K it was found to have a tensile strength of 1.3 GPa and a tensile ductility of 90%, with a crack-initiation toughness of $K_{JIc}$ = 273 MPa√m and crack-growth $K_{ss}$ toughness exceeding 400 MPa√m.[14] Although strength and toughness are often mutually exclusive properties,[15] the CrCoNi alloy exhibits exceptionally high damage tolerance with fracture toughness values amongst the largest ever reported. Such CrCoNi-based multiple principal element alloys are clearly strong candidate materials for potential applications in extreme environments, such as at very high strain rates and cryogenic temperatures.

In light of their remarkable damage-tolerance, we investigate here their fracture toughness at even lower temperatures (~20 K) by performing nonlinear-elastic *J*-based fracture toughness tests[16] in a liquid helium environment. Impact tests on the CrCoNi alloy have been reported to show high Charpy V-notch energies of close to 400 J at 77 K which were reduced by approximately 10% at 4.2 K.[17] However, despite their exceptional cryogenic mechanical properties, it remains unclear how samples that contain a sharp crack would perform at temperatures below 77 K where anomalies in the temperature-dependence of strength and ductility have been reported.[18-21] Furthermore, full resistance-curve measurements that define both the crack-initiation and crack-growth fracture toughness have not been performed on medium-/high-entropy alloys at low temperatures approaching that of liquid helium. Here, in addition to measurements of the $J_{Ic}$ initiation toughness and $J_{ss}$ growth toughness (and their corresponding stress-intensity-based values), we perform *in situ* neutron diffraction measurements and extensive post-fracture electron backscatter diffraction (EBSD) analysis, fractography, and particularly transmission electron microscopy to examine in detail the salient plastic deformation mechanisms and defect behavior that represent the fundamental basis of their exceptional fracture resistance which we find progressively increases with decreasing temperature, contrary to most metallic materials.

The CrCoNi and CrMnFeCoNi alloys investigated here were arc melted, drop cast, and homogenized at 1200°C prior to being cold worked at room temperature and recrystallized at 800°C to give a single-phase equiaxed grain structure with an average grain size of ~21 μm in CrMnFeCoNi and ~8 μm in CrCoNi; this is described in the Materials and Methods section of the Supplementary Materials. Based on the experimental procedures also detailed in that section, the experimentally measured crack-resistance curves (R-curves) of these two alloys at 20 K are shown in Figs. 1A and B; the data are compared with R-curves taken at ambient (293 K), dry ice (198 K) and liquid nitrogen (77 K) temperatures. Corresponding stress-intensity based fracture toughness values, back-calculated from the *J*-values, are shown in Figs. 1C and D as a function of temperature (20 K to 293 K). Plotted are the crack-initiation toughness, $K_{JIc}$, determined according to ASTM Standard E1820,[16] and the crack-growth toughness, $K_{ss}$, defined at the ASTM E1820 maximum limit of valid crack extension[16] where $\Delta a$ = 2.25 mm (near the plateau of the R-curve).

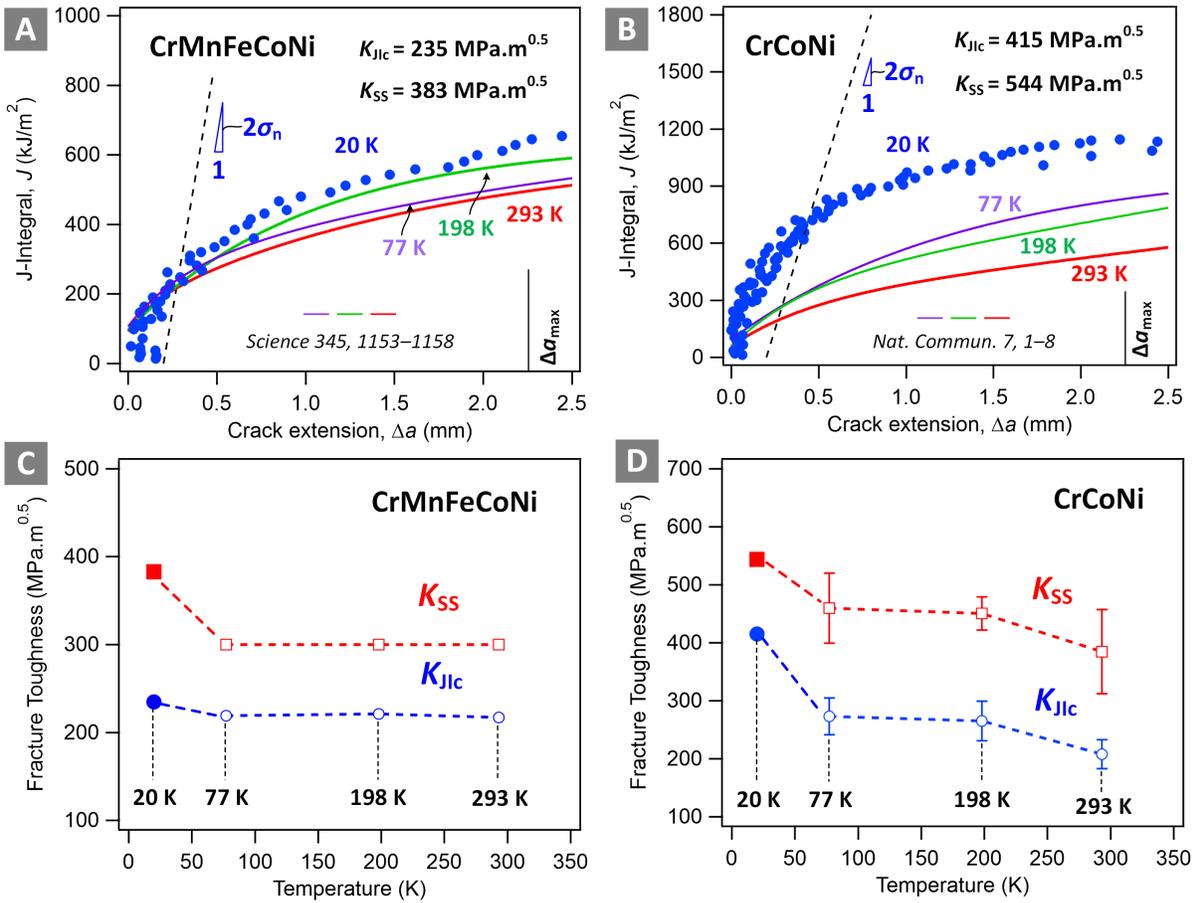

**Fig. 1. *J*-R curves and fracture toughness values for the CrCoNi and CrMnFeCoNi alloys as a function of temperature.** *J*-R curves showing the variation in the *J*-integral as a function of crack extension $\Delta a$ for (A) the CrMnFeCoNi HEA and (B) the CrCoNi MEA, between room temperature (RT ~ 293 K) and close to liquid-helium temperatures (20 K). Corresponding *K*-based fracture toughness values for (C) the CrMnFeCoNi HEA and (D) the CrCoNi MEA back-calculated from these R-curves, where $K_{JIc}$ represents the crack-initiation toughness and $K_{ss}$ the crack-growth toughness, defined at the ASTM E1820 maximum limit of valid crack extension where $\Delta a$ = 2.25 mm. Note how the toughness of both alloys at 20 K is higher than at other temperatures. The $K_{JIc}$ and $K_{ss}$ values for the CrCoNi alloy are believed to be amongst the highest toughnesses ever reported.

Both alloys show markedly rising R-curves which progressively increase with decreasing temperature, especially CrCoNi. Both CrMnFeCoNi and CrCoNi exhibit especially high fracture toughness values near liquid helium temperatures; the crack-initiation $K_{JIc}$ and crack-growth $K_{ss}$ values for the CrMnFeCoNi alloy are, respectively, 235 MPa√m and 383 MPa√m at 20 K, whereas the corresponding values for CrCoNi are 415 MPa√m and 544 MPa√m. The latter exceed the highest toughnesses that we are aware of. Fracture surfaces at 20 K showed no sign of any brittle fracture features and exhibited 100% ductile failure by microvoid coalescence, similar to earlier work at higher temperatures of 77 K to 293 K,[10,14] with dimple sizes in the range of several micrometers (Fig. 2).

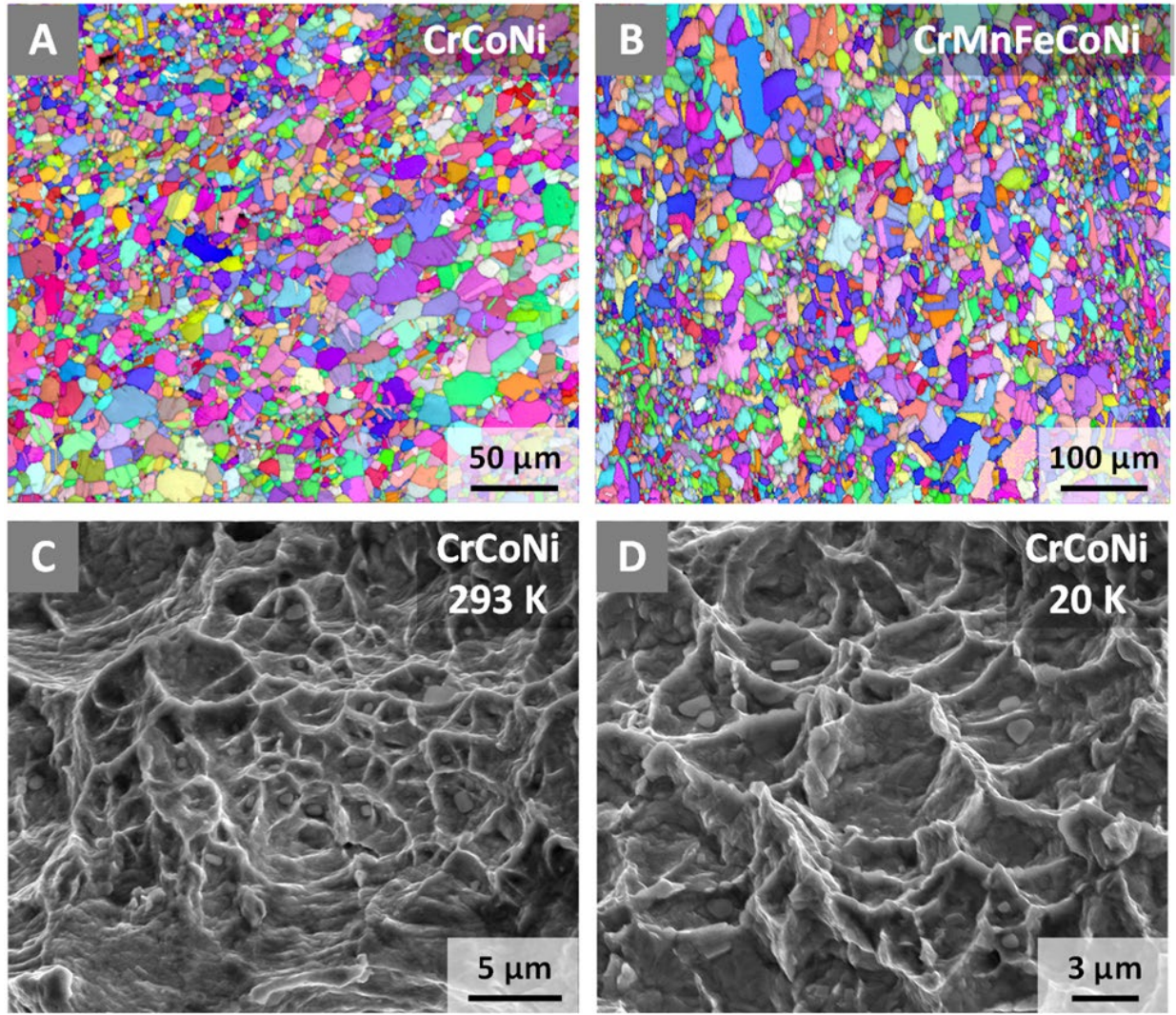

**Fig. 2. Microstructure and fractography of the CrCoNi-based alloys.** Equiaxed single-phase microstructures of (A) CrCoNi and (B) CrMnFeCoNi alloys. Fracture in both alloys at temperatures between ambient and liquid helium temperatures occurs by microvoid coalescence. Such ductile fractures in CrCoNi are shown at (C) 293 K and (D) 20 K.

It is important to note here that, despite their exceptionally high fracture toughness, these alloys do not have complex microstructures - they are simple single-phase solid solutions (Figs. 2A,B). Thus, the question that immediately arises is the origin of this remarkable fracture resistance, and why it should be so progressively enhanced at cryogenic temperatures.

To address this, we look to the cooperative defect behavior responsible for plastic deformation in these alloys,[22-24] using mainly the CrCoNi alloy to illustrate the prototypical behavior at 20 K as compared to room temperature. We employ post-fracture EBSD analysis and high-resolution transmission electron microscopy (HRTEM) of the heavily deformed regions within the plastic zone, directly adjacent to the crack tip where local strains can readily be on the order of 60 to 100%. As we describe below, while the microstructure starts off as a rather simple single-phase solid solution, deformation at these extremely low temperatures transforms the structure into an incredibly rich and complex mixture of phases and defect structures.

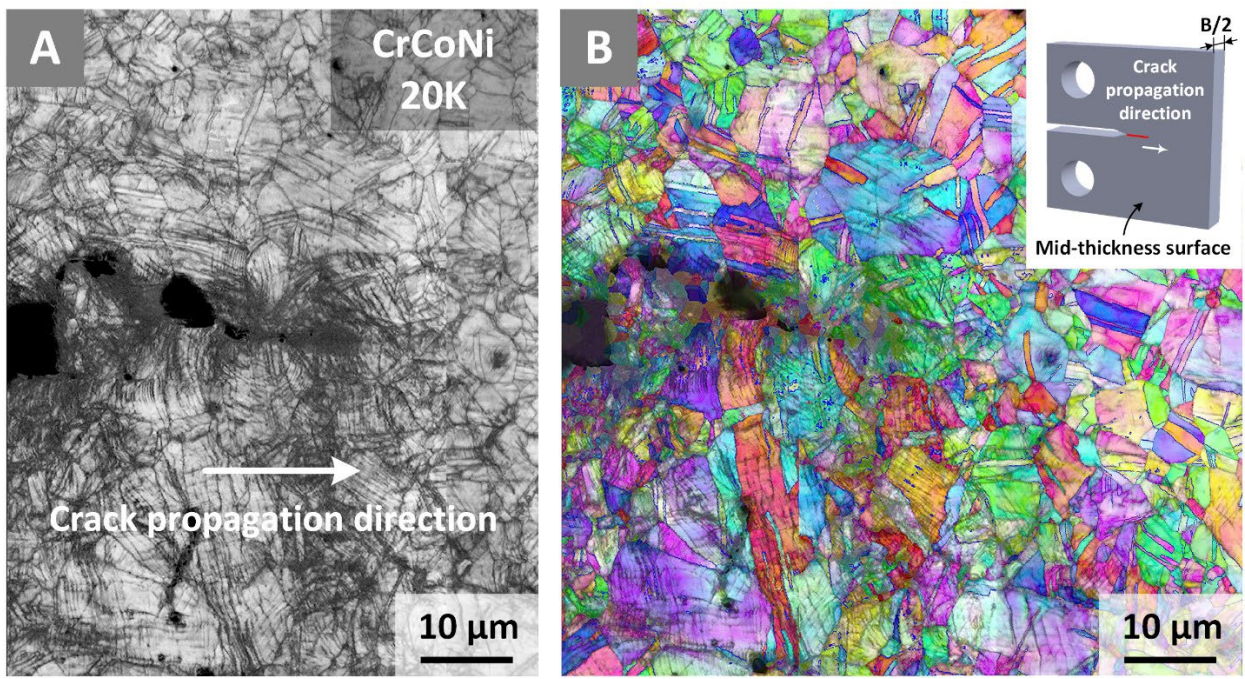

**Fig. 3. Electron backscatter diffraction (EBSD) (A) image quality (IQ) map and (B) inverse pole figure (IPF) map showing the fracture path and accompanying deformation behavior of the CrCoNi alloy at 20 K.** For the fracture propagating from left to right, predominantly plane-strain sections (taken at mid-thickness of the compact-tension samples) show the microstructure in the heavily deformed region directly ahead of the crack tip (within the plastic zone). Profuse deformation nano-twinning and stacking fault formation is readily apparent, as indicated in the EBSD images.

To investigate the microscopic deformation mechanisms, the C(T) samples used in the fracture toughness tests were first cut through the mid-thickness to obtain sections of the fracture path that were predominantly under plane-strain conditions. These were mechanically and then electrolytically polished for examination with EBSD. EBSD image quality (IQ) and inverse pole figure (IPF) maps, shown in Fig. 3 for the CrCoNi alloy tested at 20 K, indicate extensive deformation-induced nano-twinning in the highly deformed grains (under high triaxiality) within the plastic zone in the vicinity of the crack-tip. Similar deformation mechanisms have been reported for CrCoNi-based alloys at these low temperature in uniaxial tensile tests where the degree of triaxiality is far lower.[18,19,21,24] Sections from these regions were then made into TEM foils using a focused ion beam (FIB) lift-out method finishing with a 5kV Ga+ polish for HRTEM, and scanning transmission electron microscopy (4D-STEM). As illustrated in Figs. 4A-I, the CrCoNi samples deformed and fractured at both room temperature and at 20 K show a strong propensity of planar deformation features. HR-TEM imaging was conducted to identify the microstructure of the planar features. As summarized in Figs. 4A,B, the planar deformation features at room temperature include both nano-twins and bands of stacking faults, but no well-defined sequence (exceeding 3 layers) of any *hcp* phases could be identified. It is worth noting that a considerable number of dislocations can be identified in the area between the planar features; this is shown in detail in fig. S1 in the Supplementary Material. On the other hand, samples tested at 20 K demonstrated a decreased tendency for nano-twinning, as well as a smaller size of the nano-twins compared to those formed at room temperature. At these very low temperatures, the dominant planar features are deformation bands full of stacking faults (Fig. 4D), with the frequent appearance of "laths" of the *hcp* phase with a thickness of a few nanometers (Fig. 4H).

Successive four-dimensional scanning transmission electron microscopy (4D-STEM) experiments were conducted to identify the size and distribution of these microstructural features. Both virtual dark-field images and selected-area diffraction patterns were reconstructed to extract the spatially-resolved structural information of the planar features in samples tested at 293 K and 20 K. As shown in Fig. 4C, the planar features generated at room temperature are primarily well-defined nano-twins or stacking faults with the former having sizes in the range of several nanometers. In contrast, the planar deformation features identified at 20 K contain a combination of diffuse yet finer nano-twins, stacking faults and well-defined hexagonal closed packed (*hcp*) phase, as shown in Figs. 4F,I. We believe that this change in deformation modes at liquid helium temperatures, promoted by the very low stacking-fault energies in these alloys[*], is primarily responsible for the ground-breaking fracture toughness.

The low deformation temperature limits dislocation motion and twin growth by suppressing thermally-activated processes. As a result, the increased flow stress increases the heterogeneous nucleation of twins and *hcp* phases at 20 K, as compared to room temperature where no *hcp* formation was detected. As twinning in these alloys has been shown to be stress-controlled,[31,32] an *fcc*→*hcp* transformation (which involves a similar change in stacking sequence) would be expected to be favored as the flow stress increases with decreasing temperature. A diffuse network of the resulting planar deformation features – nano-twin and phase interfaces - acts to further decrease the mean free path for dislocation motion. Combined with suppressed dynamic recovery at these low temperatures, a synergy of deformation mechanisms – dislocation glide, stacking fault formation, deformation nano-twinning, transformation-induced plasticity (TRIP), and at very low temperatures inhomogeneous deformation evidenced by serrations in the stress-strain curves (as reported in refs. 17-21) - is created at increasing strain levels, which presents a highly efficient process for developing and most importantly prolonging strain hardening to restrict the localization of deformation in the crack-tip region. In simple terms, the strain hardening naturally increases strength but at the same time delays the onset of necking which promotes ductility; the corresponding elevation in strength and ductility resulting from these multiple deformation mechanisms thereby enhances the toughness.

---

[*] Stacking-fault energies, respectively for CrMnFeCoNi[25] and CrCoNi[26], have been experimentally determined as ~30 and 14 mJ/m$^2$ at room temperature, which further progressively decrease at lower temperatures. Indeed, measurements in CrFeCoNiMo$_{0.2}$ report the stacking-fault energy to decrease from 28 mJ/m$^2$ at 293 K to 11 mJ/m$^2$ at 15 K[27]. However, microscopy-based measurements obtained from balancing forces on finitely dissociated dislocations in concentrated alloys have drawn criticism for neglecting local variation in the Peierls potential due to chemical fluctuations[28] and their associated lattice distortion[29], as well as grain-size dependence.[30] Nevertheless, the trend of reduced stacking-fault energies at lower temperatures in these alloys is inarguable and consistent with theoretical predictions of the increasing energetic stability of the *hcp* phase relative to the *fcc* phase with decreasing temperature.

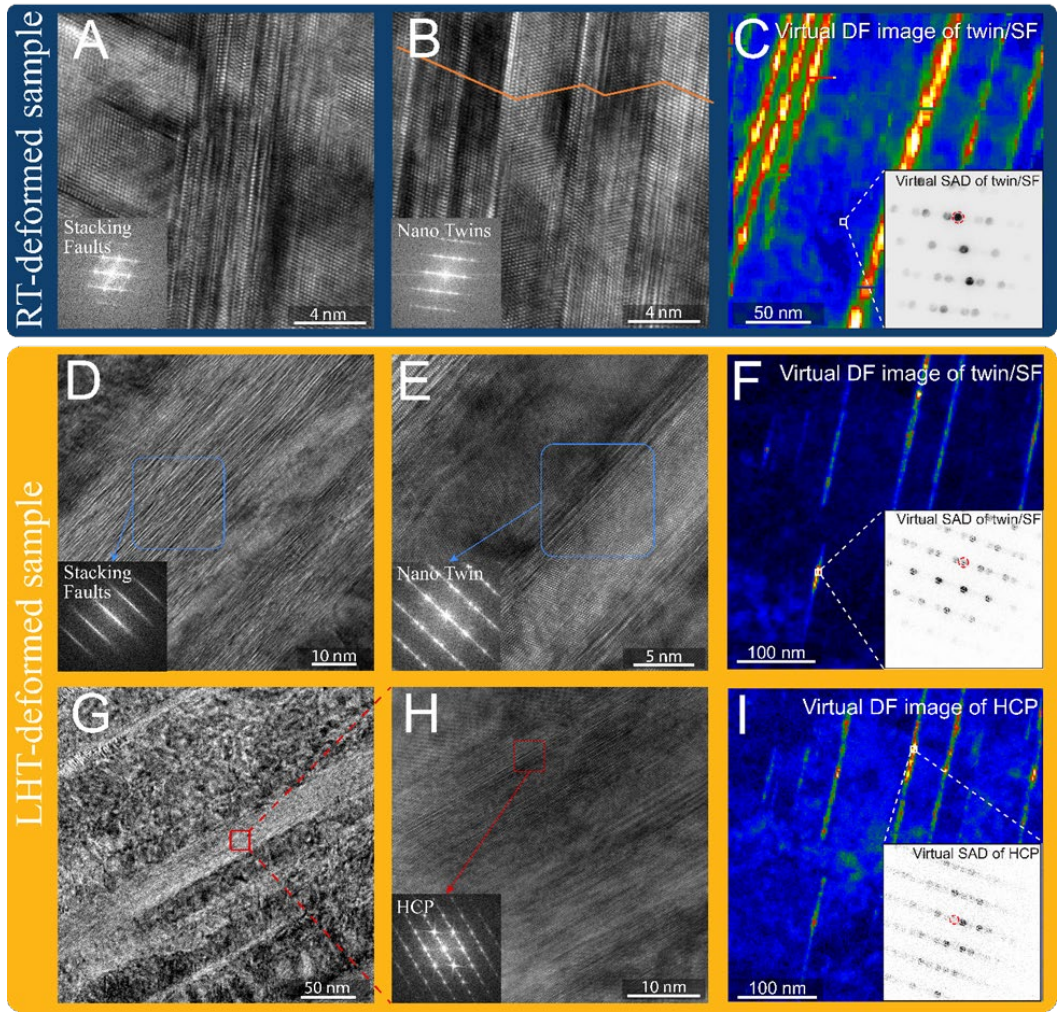

**Fig. 4. HRTEM and 4D-STEM characterization of the deformed microstructures adjacent to the fracture surfaces at 293 K and 20 K.** (A)-(C) and (D)-(I), Overview of STEM and bright-field images of samples tested at 293 K (RT-deformed sample) and at 20 K (liquid helium temperature (LHT)-deformed sample), respectively. (A) and (B) HRTEM images showing representative deformation bands with stacking faults and nano-twins, respectively. (C) Virtual dark-field image generated from the 4D-STEM scan (stronger intensity is in warmer color). The position of the virtual aperture is marked by the red circle in the inset. The inset is a virtual selected-area diffraction pattern generated from the region marked in (C), showing the diffraction from the nano-twins. (D) HRTEM image showing a representative deformation band in the sample. As shown in the inset, a high density of stacking faults can be identified in the band. (E) A nano-twin identified in the 20 K sample. (F) Virtual dark-field image generated from the 4D-STEM scan. The virtual aperture is set to pick up the signal from twins and stacking faults, as marked by the red circle in the inset. (G) Bright-field image of deformation bands (H) HRTEM image taken from a [110] orientation at 20 K. Multiple nm-sized bands in *hcp* sequence can be identified. The inset shows the Fast Fourier Transform (FFT) image from one of the *hcp* bands, representing a $[11\bar{2}0]$ orientation. (I) Virtual dark-field image generated from the 4D-STEM scan. The virtual aperture is set to pick up the signal from the *hcp* phase, as marked by the red circle in the inset. The insets in (F) and (I) are virtual diffraction patterns from the twin/stacking faults, and the *hcp* phase, respectively.

The role of the TRIP effect is particularly interesting here. The *in situ* transformation to the hexagonal phase has now been reported in numerous papers.[33-35] It provides for further strain hardening at higher strains but the resulting epsilon-martensite is not necessarily beneficial for toughness, as *hcp* phases are generally not as ductile as *fcc* phases, which in significant amounts would likely cause a ductile-brittle transition at temperatures as low as liquid helium temperatures. Indeed, some authors have reported a decrease in ductility at 4.2 – 20 K, as compared to 77 K, in similar alloys which they attribute to ε-martensite formation.[19,36] Intriguingly, 0 K *ab initio* calculations suggest that, for a given degree of (dis)order, *hcp* CrCoNi is lower energy than the *fcc* equivalent; this relationship inverts at higher temperatures due to the activation of phonon modes[37] and also likely spin fluctuations.[38] This picture is consistent with the observation of somewhat larger *hcp* lamellae at 20 K, but it must be asked why, during conventional low-temperature deformation, the *hcp* phase is still only found in quantities ranging from 0 to a maximum of a few vol.%.[17,24,39] This is also consistent with our *in situ* neutron diffraction data (fig. S5) where the maximum stacking fault probability caused by deformation was estimated to be $6.3 \times 10^{-3}$. One particularly enticing explanation for the observed behavior is the presence of quenched-in chemical short-range order (SRO) stabilizing the *fcc* matrix relative to the formation of less-ordered *hcp* regions.[39] Recent simulations[41] offer specific insight to the mechanisms by which *hcp* and twin nuclei expand in random, but not in short-range ordered samples at room temperature. While the material used in this study was not prepared in a manner intended to promote local ordering,[42] EXAFS measurements[25] and diffuse intensity in diffraction data[43] have previously indicated the presence of short-range order in water-quenched samples, and it is conceivable that conventionally processed CrCoNi contains some degree of elusive atomic-scale chemical ordering. In this case, the physics of deformation would be similar to those discussed by Yu *et al*.[41], with a greater driving force for *hcp* formation at lower temperatures. Whatever the reason, because the *hcp* laths are relatively thin, they can provide a disproportionally large number of barriers to dislocation motion, even when the overall volume fraction is small, as in the case of nanotwins[31]. For example, a single *hcp* lath with finite thickness introduces two new barriers within a grain, thereby splitting the grain into two or three segments (depending on lath thickness). For the sake of simplicity, if the *hcp* lath is assumed to be very thin and its boundaries are assumed to have the same strength as grain boundaries, the effective grain size is halved when one lath is introduced in each grain, quartered when three laths are introduced per grain, and so on. This would induce significant strengthening by a Hall-Petch type mechanism in the strain-hardening regime because the phase transformation occurs dynamically during straining. As is the case with nanotwins in a polycrystalline HEA[31], the main role of the *hcp* phase is to enhance strain-hardening, which leads indirectly to increased ductility by postponing necking instability rather than directly by accommodating the applied strain. Despite the *hcp* phase being brittle, this nevertheless serves to enhance the toughness, but with the key feature that its volume fraction must be small.

We conclude that as cryogenic structural materials, equiatomic, single-phase *fcc* CrCoNi-based medium- and high-entropy alloys, in particular the CrCoNi alloy, possess some of the most impressive mechanical properties of any metallic alloy reported to date. Indeed, their crack-initiation and crack-growth fracture toughness values at 20 K are among the highest ever recorded (Fig. 5), a fact that we ascribe to their effective strain hardening capacity generated by their synergy of deformation mechanisms created under increasing strains, including stacking fault formation, deformation nano-twinning, and a TRIP process which results in limited *hcp* ε-martensite formation.

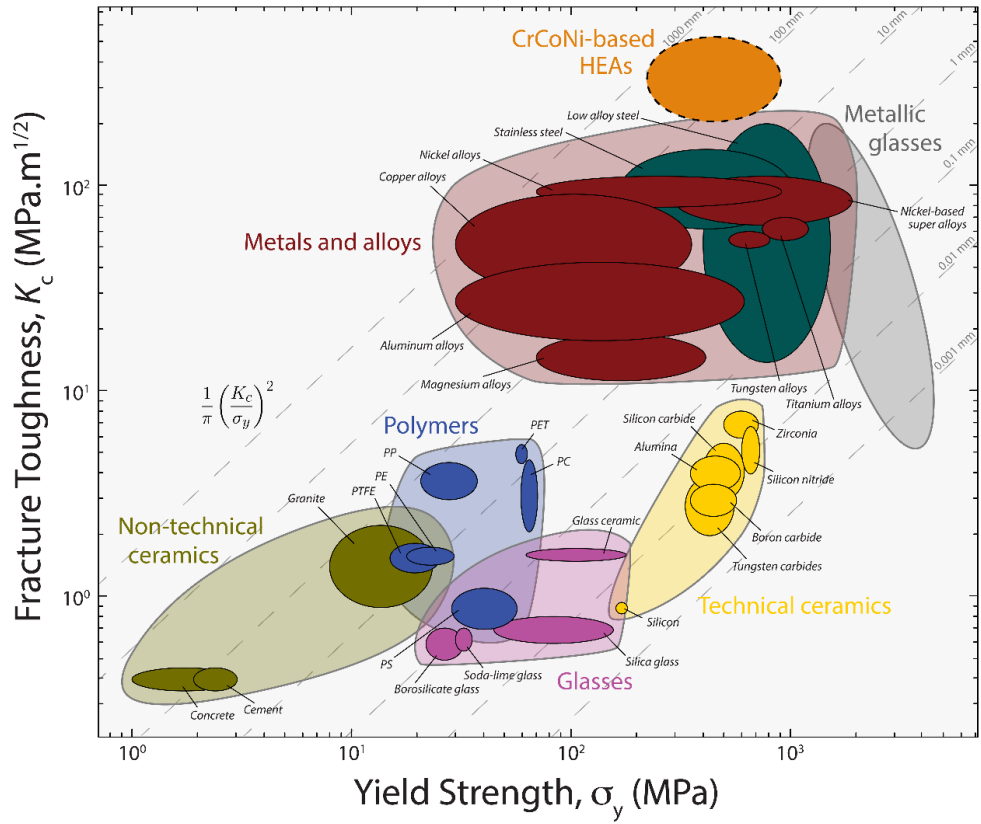

**Fig. 5. Ashby map in terms of the fracture toughness, $K_c$, versus the yield strength, $\sigma_y$, for a broad class of materials.** Note the remarkable fracture toughness of the CrCoNi-based medium- and high-entropy alloys at tensile strengths of ~1 GPa, which appear to be the highest on record.

**ACKNOWLEDGMENTS**

**Funding:** The research was primarily supported by the U.S. Department of Energy, Office of Science, Basic Energy Sciences, Materials Sciences and Engineering Division, through the Damage-Tolerance in Structural Materials program (KC13) at the Lawrence Berkeley National Laboratory (LBNL) under contract no. DE-AC02-CH11231, and the Multiscale Mechanical Properties and Alloy Design program (ERKCM06) at the Oak Ridge National Laboratory. The authors acknowledge the use of the ENGIN-X, ISIS Facility, at the Rutherford Appleton Laboratory, Chilton, Didcot, Oxon, U.K., for the mechanical testing in liquid helium, and the microscopy facilities in the National Center for Electron Microscopy, in the Molecular Foundry at LBNL, which is supported by the Office of Science, Office of Basic Energy Sciences of the U.S. Department of Energy under contract DE-AC02-05CH11231. D.L., T.J. and P.F.-K. acknowledge support from the U.K. Engineering and Physical Sciences Research Council (grants # EP/N004493/2, EP/T000368/1), and B.G. from the ARC Future Fellowship (project FT190100484) and the UNSW Scientia Fellowship schemes.

**Author contributions:** R.O.R., B.G., and E.P.G. formulated the original idea. D.L. and R.O.R secure instrumental time at ENGIN-X via a beamline proposals. Using materials provided by E.P.G., D.L. P.F.-K. and S.K. performed the toughness tests, which were analyzed by Q.Y., B.G. and D.L. under the supervision of R.O.R. Neutron diffraction data were analyzed by M.J. and D.L. Structural characterization was performed by Q.Y. (EBSD) and R.Z and M.P. (TEM), the latter under the supervision of A.M.M.  Simulation studies were conducted by F.W. under the supervision of M.A.  R.O.R. drafted the manuscript, with all authors contributing. R.O.R. supervised the project.

**Competing interests:** The authors declare no competing interests, financial or otherwise.  Readers are welcome to comment on the online version of the paper.

**Data and materials availability:** All data that support the findings of this study are reported in the main paper and supplementary materials.  Any correspondence should be addressed to roritchie@lbl.gov (R.O.R.).

**SUPPLEMENTARY MATERIALS**

Materials and Methods
Figs. S1-S5

# Supplementary Materials for

## Exceptional fracture toughness of CrCoNi-based medium- and high-entropy alloys close to liquid helium temperatures

**Dong Liu, Qin Yu, Saurabh Kabra, Ming Jiang, Paul Forna-Kreutzer, Ruopeng Zhang, Madelyn Payne, Bernd Gludovatz, Mark Asta, Andrew M. Minor, Easo P. George, Robert O. Ritchie**

Correspondence to: roritchie@lbl.gov (R.O.R.)

**This PDF file includes:**

Materials and Methods
Figs. S1-S5

## Materials and Methods

<u>Materials Synthesis and Processing</u>

Following the process described in earlier papers,[1] equiatomic CrMnFeCoNi and CrCoNi alloys were produced from high-purity elemental starting materials (>99.9% pure) by arc-melting and drop-casting in an argon atmosphere into rectangular cross-section copper molds measuring approximately 25.4 mm by 19.1 mm by 127 mm. Castings of both alloys were homogenized at 1200°C for 24 h *in vacuo*, cut in half length wise and then cold forged and cross-rolled at room temperature to achieve plates with a thickness of ~10 mm. These sections were annealed in air at 800°C for 1 h to generate fully recrystallized, equiaxed grain structure with an average grain size of ~21 μm in CrMnFeCoNi and ~8 μm in CrCoNi.

<u>Sample Preparation</u>

Compact-tension (C(T)) samples were machined from the recrystallized plates. For the CrCoNi alloy, C(T) samples of width, $W$ = 18 mm and thickness, $B$ = 5 mm, were prepared by electrical discharge machining (EDM) in accordance with ASTM standard E1820.[2] For the CrMnFeCoNi alloy, the same procedures were adopted except that the sample thickness, $B$, was 9 mm. In both types of samples, notches ~6.3 mm in length with a root radius of ~100 μm were introduced by EDM. Pre-cracking by fatigue was performed on all samples at room temperature using a servo hydraulic MTS 810 load frame (MTS Corporation, Eden Prairie, MN, USA) with an Instron 8800 digital controller (Instron Corp., Norwood, MA, USA). These fatigue cracks were introduced by tension-tension loading under load control at a $\Delta K$ of ~15 MPa√m, at a load ratio, $R$ = 0.1 at a constant frequency of 10 Hz (sine wave). Crack extension was monitored by mounting an Epsilon clip gauge (Epsilon Technology, Jackson, WY, USA) with a 3 mm (-1/+2.5 mm) gauge length, at the load line of the specimen. Optical microscopy was used to inspect the crack length from both sides of the sample to ensure a straight crack front. The final crack lengths were no less than 9 mm in length with an approximate $a/W$ of 0.5 to ensure a pre-crack length well above the minimum length (~1.3 mm) required by ASTM E1820.[2]

Experimental Mechanical Testing Setup

Fracture toughness measurements in both alloys were performed at ENGIN-X, the dedicated materials engineering beamline at the ISIS Neutron and Muon Source at the Rutherford Appleton Laboratory at Harwell campus, Oxfordshire, in the U.K. Nonlinear elastic fracture mechanics methodologies were employed to determine the fracture toughness in terms of crack-resistance $J_R$ curve, representing the change in crack resistance with increasing crack extension in the pre-cracked C(T) samples. These samples were loaded under displacement control at ~2 mm/min with a periodic unloading sequence (unloading compliance) adopted to determine the crack extension, $\Delta a$. Unloading was controlled by a Python script in which the load was reduced by ~10% at each unloading cycle, and the unloading compliance measured using the Epsilon clip gauge.

At ENGIN-X, a unique cryogenic testing chamber was used, the details of which are described elsewhere.[3,4] Briefly, it consists of a vacuum vessel with aluminium windows to permit access for the incident and scattered neutrons (fig. S1). Two cryo-coolers were connected to the infrared radiation shield of the chamber to control the temperature. The load, applied by an Instron 100 kN rated hydraulic frame with an Instron 8800 Controller, was transmitted to the sample by a shaft made from Ferralium 255 SD50 material and connected to a threaded section made from Inconel 718. Inside the chamber, the Inconel 718 shaft pulls into the CuBe pivot which is insulated by G10 cone and spacer (dielectric material with low thermal conductivity and high strength). A CuBe adaptor shaft is locked in position with CuBe nuts on a link plate that is connected directly to the first stage of the cryogenic cooler; a schematic of the grips assembly is shown in fig. S2. The sample holder grips are thermally linked through a copper clamp and a stack of laminated copper sheets for better conductivity.

A set of CuBe sample holder with CuBe pins were designed and installed to the cold fingers (fig. S3a). Two rhodium-iron temperature sensors: one on the CuBe cold fingers and the other one screwed to the bottom of the sample away from the crack tip (fig. 3b). The testing of the samples was conducted once the temperature sensor readings stabilised around ~10-15 K at the cold fingers and ~25-30 K at the bottom of the samples. Therefore, it is reasonable to take the average temperature from these temperature sensor readings as the temperature at the crack tip, namely, around 17.5-22.5 K. We approximate this as 20 K in the text.

For both alloys, one sample of each was tested *in situ* during neutron diffraction measurement. An additional three samples of each were tested *ex situ* to repeat the experiment and one more sample of each alloy was tested for comparison at room temperature (293 K).

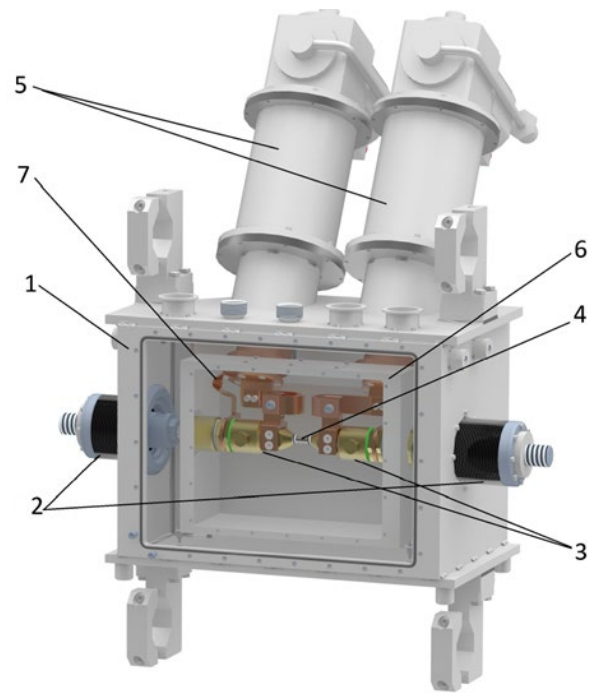

**Fig. S1**. **Schematic illustration of the ENGIN-X cryogenic testing chamber**: (1) outer vacuum vessel; (2) two bellow sections; (3) sample grips; (4) sample; (5) two cryo-coolers; (6) infrared radiation shield; (7) copper braids.[3]

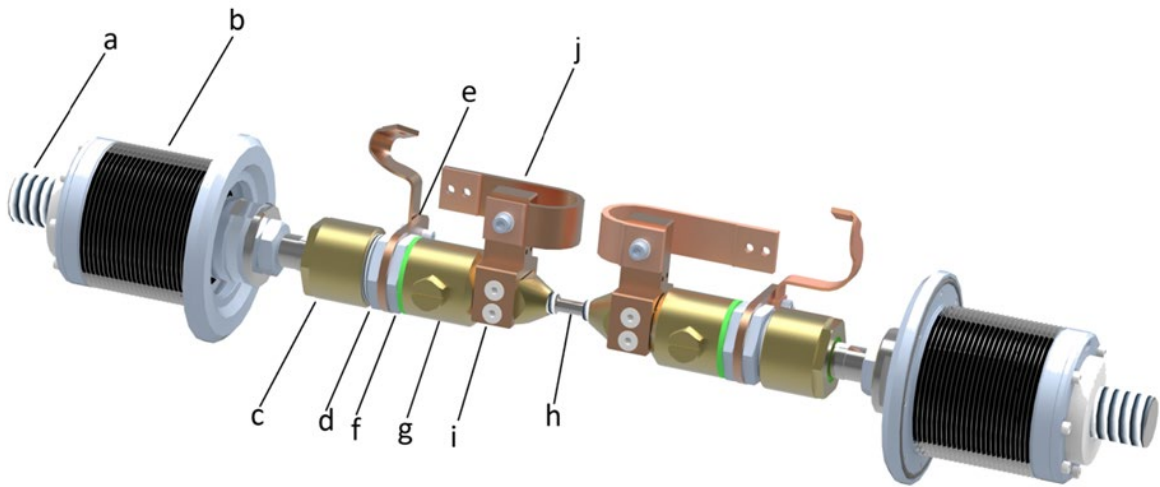

**Fig. S2. A detailed view of the grips assembly:** (a) shaft; (b) bellow section; (c) copper pivot; (d) adaptor shaft; (e) copper link; (f) G10 spacer; (g) copper-beryllium sample grip; (h) sample; (i) copper clamp; (j) stack of laminated copper sheets.

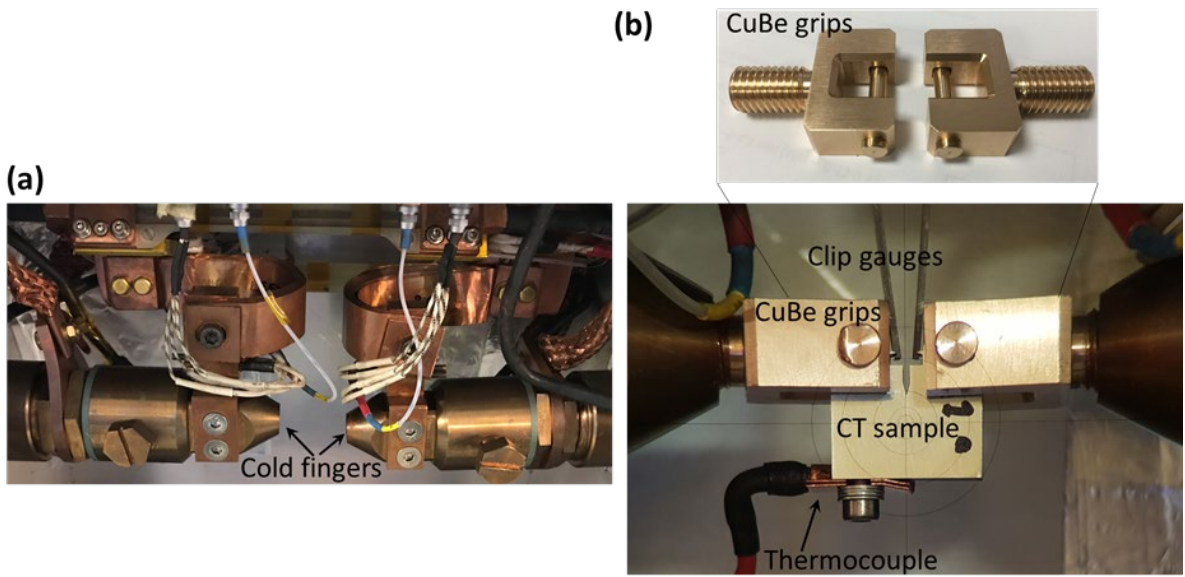

**Fig. S3. Further details of the specimen gripping assembly:** (a) A close-up view of the grips assembly. The samples are loaded between the cold fingers. (b) Photograph showing the set-up of the clip gauge, CuBe grips, the C(T) samples held by CuBe pins and a temperature sensor installed at the bottom face of the C(T) sample.

*In Situ* Neutron Diffraction

As ISIS is a time-of-flight (ToF) neutron spallation source, at the ENGINX beamline a high flux neutron beam with narrow pulse width is available with a 50 m primary flight distance ($L_1$) to the test specimen. It has two collimated detectors 180° ($2\theta$) apart at 90° and -90°, respectively, to the incident neutron beam. An instrumental gauge volume (IGV) located inside the C(T) specimen ahead of the crack tip was used to measure at about 4 x 4 x 4 mm³. The distance between the IGV and the detectors was 1.53 m (the secondary neutron flight length, $L_2$). The detected neutron wavelength, $\lambda$, and the measured flight time, $t$, are related by:

$$\lambda = \frac{h}{m(L_1+L_2)} t \quad , \tag{1}$$

where $h$ is Planck's constant, and $m$ is neutron mass. As such, the ToF spectrum can be converted to a diffraction pattern with each peak corresponding to a family of {$hkl$} planes. The spacing of the crystal lattice planes, $d_{hkl}$, can be derived by:

$$d_{hkl} = \frac{h}{2\sin\theta\, m(L_1+L_2)} t_{hkl} \quad , \tag{2}$$

where $\theta$ is the half angle between the incident and scattered neutron beams ($\theta$ = 45° was used in the current setup). Prior to the measurements, a calibration process was performed on a standard $CeO_2$ powder sample.

All the neutron spectra collected were analysed using a single-peak fitting with a Pseudo Voigt profile (*i.e.*, mixed Gaussian and Lorentzian) convoluted with a sharp-edged exponential on a linear background. The centroid of the fitted peak in *d*-spacing units was taken as the lattice spacing. One sample from each alloy was tested *in situ* with 24 neutron spectra collected for each sample, first in a nominally load-free condition (50 N pre-load) from 293 K to 20 K, and then at increasing loading steps until final fracture at 20 K. The calculated *d*-spacings and the crystal lattice shrinkage with temperature were derived and are presented below.

Despite the fact that our TEM images did reveal limited evidence of the formation of the *hcp* phase, our *in situ* neutron diffraction measurements did not. Of note here is that He *et al.*[5] reported the formation of 4 vol.% of *hcp* phase based on the (101) peak in neutron diffraction spectra collected from a uniaxial tensile specimen tested at 15 K with a strain between 14% and 30%.

However, in the present work the highly strained volume ahead of the crack tip is at least an order of magnitude smaller compared with a uniaxial tension configuration. With a 4x4x4mm instrumental gauge volume used in the neutron measurement, the small fraction of *hcp* phase most likely did not yield sufficient diffraction signal discernable in the overall spectrum.

Fracture Mechanics Testing

We used nonlinear-elastic fracture mechanics methods, which incorporate both the elastic and inelastic contributions, to measure the fracture toughness; specifically, the change in crack resistance with crack extension, *i.e.*, crack-resistance curve (*R*-curve) behavior, was characterized in terms of the *J*-integral as a function of crack growth primarily at 20 K (close to liquid helium temperature) but also at 293 K (in room air) for comparison. Prior fracture toughness measurements in these alloys in room air, dry ice and liquid nitrogen are also included in the paper for comparison; these have been described in detail elsewhere.[6,7] Corresponding measurements at liquid helium temperatures were performed at ENGIN-X, as described above.

Samples were tested under displacement control at a constant displacement rate of 2 mm/min. The onset of cracking as well as subsequent subcritical crack growth were determined by the compliance method with periodically unloading the sample (~10% of the peak-load) to record the elastic unloading compliance using a clip gauge of 3 mm (-1/+7 mm) gauge length (Epsilon Technology, Jackson, WY, USA) mounted in the load-line of the sample. Crack lengths, $a_i$ were calculated from the compliance data obtained during the test using the compliance expression of a C(T) sample at the load-line[2]:

$$\frac{a_i}{W} = 1.000196 - 4.06319u + 11.242u^2 - 106.043u^3 + 464.335u^4 - 650.677u^5 , \quad (3)$$

where

$$u = \frac{1}{[BEC_{c(i)}]^{\frac{1}{2}}+1} . \quad (4)$$

In Eq. 4, $C_{c(i)}$ is the rotation-corrected, elastic-unloading compliance and *B* is the sample thickness. Initial and final crack lengths were also verified using optical measurements. For each crack length data point, $a_i$, the corresponding $J_i$-integral was computed as the sum of elastic, $J_{el\,(i)}$, and plastic components, $J_{pl\,(i)}$, such that the $J$-integral can be expressed as follows:

$$J_i = K_i^2/E' + J_{pl\,(i)}, \tag{5}$$

where $E' = E$, the Young's modulus, in plane stress and $E/(1-\nu^2)$ in plane strain; $\nu$ is Poisson's ratio. $K_i$, the linear-elastic stress intensity corresponding to each data point on the load-displacement curve, was calculated following:

$$K_i = \frac{P_i}{(B^2W)^{\frac{1}{2}}} f(a_i/W), \tag{6}$$

where $P_i$ is the applied load at each individual data point and $f(a_i/W)$ is a geometry-dependent function of the ratio of crack length, $a_i$, to width, $W$, as listed in the ASTM standard. The plastic component of $J_i$ can be calculated from the following equation:

$$J_{pl(i)} = \left[J_{pl(i-1)} + \left(\frac{\eta_{pl(i-1)}}{b_{(i-1)}}\right)\frac{A_{pl(i)} - A_{pl(i-1)}}{B_N}\right]\left[1 - \gamma_{(i-1)}\left(\frac{a_{(i)} - a_{(i-1)}}{b_{(i-1)}}\right)\right], \tag{7}$$

where $\eta_{pl\,(i-1)} = 2 + 0.522\,b_{(i-1)}/W$ and $\gamma_{pl\,(i-1)} = 1 + 0.76\,b_{(i-1)}/W$. $A_{pl\,(i)} - A_{pl\,(i-1)}$ is the increment of plastic area underneath the load-displacement curve, and $b_i$ is the uncracked ligament width (*i.e.*, $b_i = W - a_i$). Following the above approach, the value of $J_i$ at any point along the load-displacement curve can be determined; along with the corresponding crack lengths, the $J$-$\Delta a$ resistance curve can then be constructed, where $\Delta a$ is the difference of the individual crack lengths, $a_i$, during crack growth and the initial crack length, $a$, after pre-cracking.

A provisional toughness $J_Q$ can be defined by the intersection of the resistance curve with the 0.2 mm offset/blunting line ($J = 2\,\sigma_o\Delta a$; where $\sigma_o$ is the effective flow stress). This provisional toughness $J_Q$ can be considered as a size-independent (valid) fracture toughness, $J_{Ic}$, provided the validity requirements for $J$-field dominance and plane-strain conditions are met, *i.e.*, respectively that $b_0, B > 10\,J_Q/\sigma_0$, where $b_0$ is the initial ligament length. The fracture toughness, $K_{JIc}$, expressed in terms of the stress intensity then can be obtained using the standard $J$-$K$ equivalence (mode I) relationship $K_{JIc} = (E'\,J_{Ic})^{1/2}$.

In terms of these ASTM criteria for $J$-dominance and plane strain, all fracture toughness values for both alloys at all temperatures fully conform to ASTM E1820 validity,[2] with the exception of the crack-growth toughness in the CrCoNi alloy at 20 K which does not fully meet the plane strain requirement. Specifically, for CrMnFeCoNi, where the effective flow stress σₒ was 1.5 GPa at 20 K,[8] the calculated values of the critical dimension, 10 $J_Q$ / σₒ , were 1.6 mm and 4.3 mm for the initiation and growth toughnesses, respectively. Accordingly, both crack-initiation $J_{Ic}$ ($K_{Ic}$) and crack-growth $J_{ss}$ ($K_{ss}$) toughness values satisfy the specimen size requirements at 20 K, indicating they are ASTM valid. Indeed, these criteria were also met for all toughness values measured for this alloy at all temperatures tested.

For CrCoNi, the effective flow stress σₒ was also 1.5 GPa at 20 K.[5] The calculated 10 $J_Q$/σₒ values were 4.37 mm and 7.5 mm for the initiation and growth toughnesses at this temperature, respectively. The initiation toughness $J_{Ic}$ ($K_{Ic}$) values satisfy the specimen-size requirements for J-dominance and plane strain and are thus fully ASTM E1820 valid. However, the growth toughness $J_{ss}$ ($K_{ss}$) at 20 K meets the requirement for J-dominance but does not completely meet the plane strain requirement, although all other crack-initiation and crack-growth toughnesses for this alloy are fully ASTM valid at all temperatures.

The temperature-dependent values of *E* and *ν* used in this study are from the measurements of Haglund *et al.*[9] for CrMnFeCoNi and Laplanche et al.[10] for CrCoNi. In the former study, data were obtained from 300 K down to 55 K, whereas the latter study extended from 1000 K down to 293 K. To obtain elastic constants at cryogenic temperatures outside these ranges, we fitted Varshni's[16] equation [ $c_{ij}^0(T) = c_{ij}^0 - s/(e^{\frac{t}{T}} - 1)$ ] to the experimental data in the above papers[9,10] and extrapolated to lower temperatures (in Varshni's equation $c_{ij}^0(T)$ and $c_{ij}^0$ are values of the elastic constants at temperature *T* and 0 K, respectively, and *s* and *t* are fitting constants). For CrMnFeCoNi,[17] *s* and *t* for *E* are 35 GPa and 416 K, while those for *G* (shear modulus) are 16 GPa and 448 K; for CrCoNi[10] *s* and *t* for *E* are 32 GPa and 362 K, and 11.9 GPa and 362 K for *G*. Assuming isotropic elasticity: $\nu = \frac{E}{2G} - 1$. In this way, values for Young's moduli, *E*, at 293 K, 198 K, 77 K and 20 K, of 202 GPa, 209 GPa, 214.5, GPa and 214 GPa and Poisson's ratios, *ν*, of 0.266, 0.263, 0.256 and 0.259 were obtained for the CrMnFeCoNi alloy. Corresponding values were obtained for the CrCoNi alloy, namely, *E* = 229 GPa, 235 GPa, 241 GPa and 241 GPa and ν = 0.31, 0.30, 0.30 and 0.287 at 293 K, 198 K, 77 K and 20 K, respectively.

Characterization of Deformation and Fracture

After testing, the broken samples were immersed in acetone to prevent oxidation and then stored *in vacuo*. Fractographic analysis was performed with scanning electron microscopy using an FEI Strata DB235 SEM (FEI Company, Portland, OR, USA) operated in the secondary electron (SE) imaging mode at 5-15 kV, to identify the salient deformation and fracture mechanisms involved in the fracture.

To examine the microstructure and nature of the deformation mechanisms in the vicinity of, and remote from, the crack tip, the fractured samples were sectioned in two, each with a thickness of ~$B$/2, so that the crack-path profile in the plane-strain region in the interior of the sample could be analyzed. For each sample, one half was embedded in conductive resin, progressively polished to a 0.05 µm surface finish, and finally vibration polished for 12 h in colloidal silica. The microstructure along the crack wake and the crack-tip region was then characterized using electron back-scattered diffraction (EBSD) in an FEI Strata DB235 SEM (FEI Company, Portland, OR, USA) operated at 20 kV using a TEAM EBSD analysis system (Ametek EDAX, Mahwah, NJ, USA) at 35 nm.

Transmission electron microscopy (TEM) was carried out to characterize the defect structures in the vicinity of the crack. Site-specific TEM foil samples were cut from the fractured samples and thinned to electron transparency using focused ion beam (FIB) milling (FEI, Helios G4 UX). Samples were trenched from regions both near the crack tip and far away from the deformed region from test specimens that were tested at both cryogenic and room temperature. High-resolution transmission electron microscopy (HRTEM) was conducted on an FEI ThemIS 60-300 STEM/TEM operated at 300 kV to investigate the deformation microstructure after fracture toughness testing at room and liquid helium temperatures. 4D-STEM imaging of samples was conducted on the double-corrected TEAM I microscope and a FEI TitanX microscope (operated at 300 kV) at the National Center for Electron Microscopy (NCEM, LBNL). A 1 mrad bullseye aperture was used with an electron probe size of approximately 1.5 nm on the TEAM1 and 2 nm on the TitanX. The scanning step size was set to 1.5 nm, accordingly.

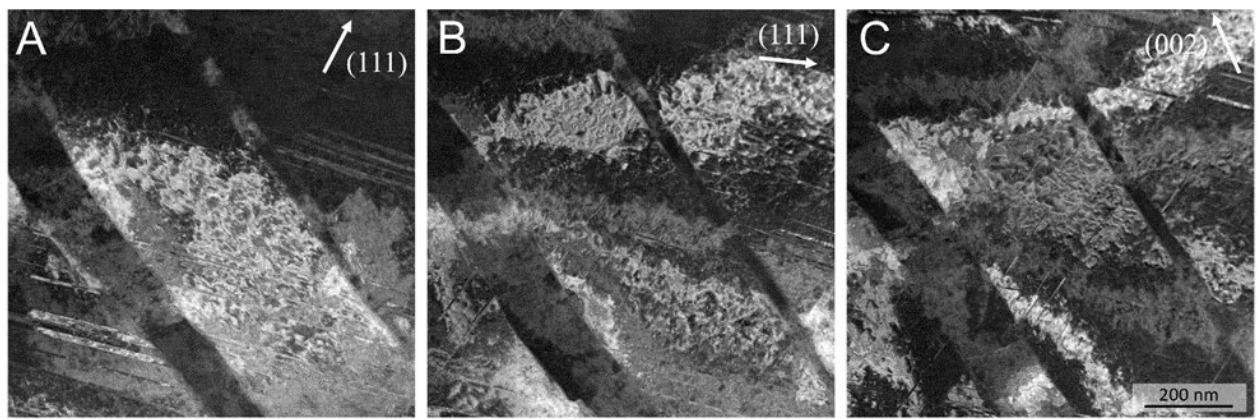

**Supplementary Fig. S4.** Diffraction contrast imaging (DCI) in STEM mode for the sample deformed and fractured at 293 K with three different two-beam conditions indicated by the white arrows. A relatively high density of dislocations can be identified in between the planar deformation bands.

Lattice parameters and peak broadening in neutron diffraction

Time-of-flight (ToF) neutron diffraction spectra collected for single phase *fcc* ternary CrCoNi material with the ToF axis being converted into interplanar *d*-spacing in Å (fig. S5a), shows the spectral evolution with cooling from 293 K down to cryogenic temperatures of 20 K and increasing the load from 10% to 75% of the maximum load. A clear peak shift in the crystal planes can be seen in both graphs as a result of cooling as well as loading. Dashed straight lines and different solid line colors in fig. S5a and b are for visual guidance only. Note that despite the fact that the TEM results in the present work did reveal limited evidence of the formation of the *hcp* phase, our *in situ* neutron diffraction measurements did not. Of note here is that He *et al.*[4] reported the formation of 4 vol.% of *hcp* phase based on the (101) peak in neutron diffraction spectra collected from a uniaxial tensile specimen tested at 15 K with a strain between 14% and 30%. However, in the present work, the highly strained volume ahead of the crack tip within the plastic zone is more than an order of magnitude smaller compared with a uniaxial tension configuration. With a 4x4x4mm instrumental gauge volume used in the neutron measurement, the small fraction of *hcp* phase did not yield a sufficient diffraction signal discernable in the overall spectrum, although the broadening of the (200) and (111) peaks reported in He et *al.*'s work[5] was detected.

The lattice constant $a_0$ for the single phase *fcc* ternary CrCoNi alloy at 293 K was determined to be 3.56 Å by fitting the {200} peak position (half of the {100}) from the diffraction spectra collected at 293 K. These data were compared with several literature values listed in fig. S5c and a good consistency can be seen. The lattice strain $\varepsilon_{hkl}$ was calculated based on the *{hkl}* planes interplanar spacing $d_{hkl}$ fitted by the Pseudo Voigt profile in a dedicated ENGIN-X's Script Based Analysis programme EX-SBA, by following the relationship:

$$\varepsilon_{hkl} \frac{(d^i_{hkl}-d^0_{hkl})}{d^0_{hkl}} \quad . \tag{8}$$

Here $d^i_{hkl}$ is the lattice spacing of *{hkl}* plane when material is stressed at each load step (*i*) where the ToF neutron diffraction scan was acquired and $d^0_{hkl}$ is the value for the stress-free material. Further, the polycrystalline coefficient of thermal expansion (CTE) $\alpha_{hkl}$ for the *fcc* CrCoNi alloy in specific *{hkl}* crystallographic planes due to thermal shrinkage at each cooling step was calculated from:

$$\alpha_{hkl} = \frac{\varepsilon_{hkl}}{\Delta T^i} = \frac{(d^i_{hkl}-d^0_{hkl})}{d^0_{hkl}\times(T^i-T^0)} \tag{9}$$

where $\Delta T^i$ is the temperature difference between each cooling temperature step ($T^i$) and initial reference temperature ($T^0$).

As shown in fig. S5d, cooling from room temperature to 20 K introduced increasing strains to the crystal lattice; the values derived for {200}, {400}, {111} and {222} planes are similar following an exponential increase with a plateau towards a lower temperature. The coefficient of thermal expansion (CTE) was derived and is shown in fig. S5e. The CTE at room temperature was determined to be 22.9±1.8 × 10$^{-6}$/K for planes in the direction of <200> and 19.8±2.0 ×10$^{-6}$/K for the <111> direction. These values are in a similar range as the average CTE values for polycrystalline CrCoNi measured by Laplanche *et al.* ~13×10$^{-6}$/K [10], and Moravcik *et al.* 17.4×10$^{-6}$/K [11]. The CTE at 20 K has been determined to be 8.0±0.2 ×10$^{-6}$/K for <200> direction and 7.8±0.3 ×10$^{-6}$/K for <111> direction; there are no open published data in the literature to compare at this temperature. Lattice strains at 20 K were further increased under increasing load. However, the load-induced lattice strains were different even on the same family of planes, *e.g.*, the {200} and {400} planes showed different strains. This is consistent with that reported by Wang *et al.*[4]. and Naeem *et al.*[8]. Since {200} and {400} planes are equivalent plane families, as are {111} and {222}, differences in their lattice strains in the same family due to loading indicates additional potential straining introduced by stacking faults as described by Wang et al[4] and Naeem et al[8]. Note that there is no difference in their strains due to cooling (fig. S5d), indicating that temperature change alone did not introduce stacking faults. Using the methodology adopted in refs. 4 and 8, the highest stacking-fault probability (SFP) in the current CrCoNi material under load at 20 K was found to be about $6.3\times10^{-3}$; we could not assign an exact strain value to this SFP as the true stress in the instrumental gauge volume ahead of the crack tip is non-uniform varying from high values in the small volume ahead of the crack tip to low levels further away from the crack.

Lastly, as mentioned earlier, peak broadening was observed, as shown by the Pseudo Voigt profile fitted Full Width at Half Maximum (FWHM) of the diffraction peaks (200) and (111) planes in the CrCoNi alloy when plotted as a function of the applied load. Consistent with the lattice strain measurement, the insert in fig. S5f shows no evidence of any thermal cooling induced peak broadening in (200) and (111). However, at 20 K when load was applied, the FWHM of the (200) and (111) planes in the CrCoNi alloy demonstrated a gradual increase with increasing load, indicating the introduction of defects as predicted from the lattice strain measurements; specifically, peak broadening of up to 60% was apparent when the load reached ~80% of the maximum load (fig. S5h). The same trend was observed in the CrMnFeCoNi alloy.

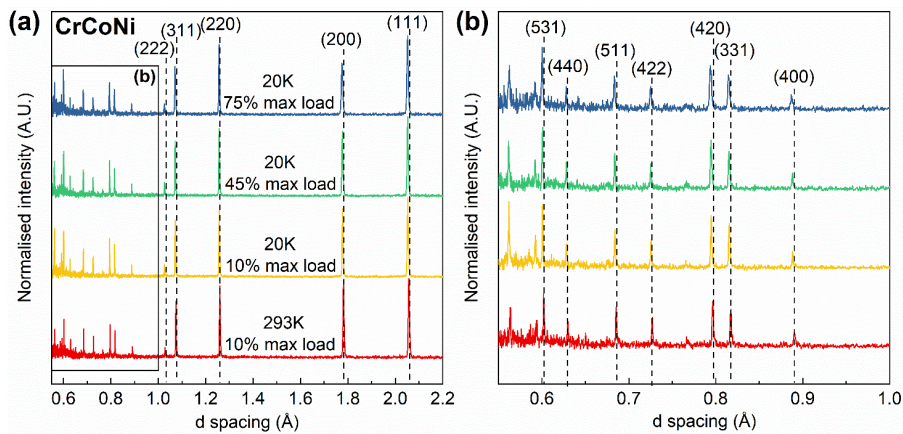

**(c)**

| Lattice constant $a_0$ (Å) at RT | Method | Authors |
|---|---|---|
| 3.559 | Pair distribution function (PDF) | Zhang et al, 2017[12] |
| 3.563 | DFT computation | Yin et al, 2020[13] |
| 3.565 | XRD | Moravcik et al, 2017[12] |
| 3.56 | XRD | Lee et al, 2019[14] |
| 3.567 | XRD | Laplanche et al, 2017[15] |
| 3.57 | XRD | Laplanche et al, 2018[10] |
| **3.562** | **neutron diffraction** | **Present work** |

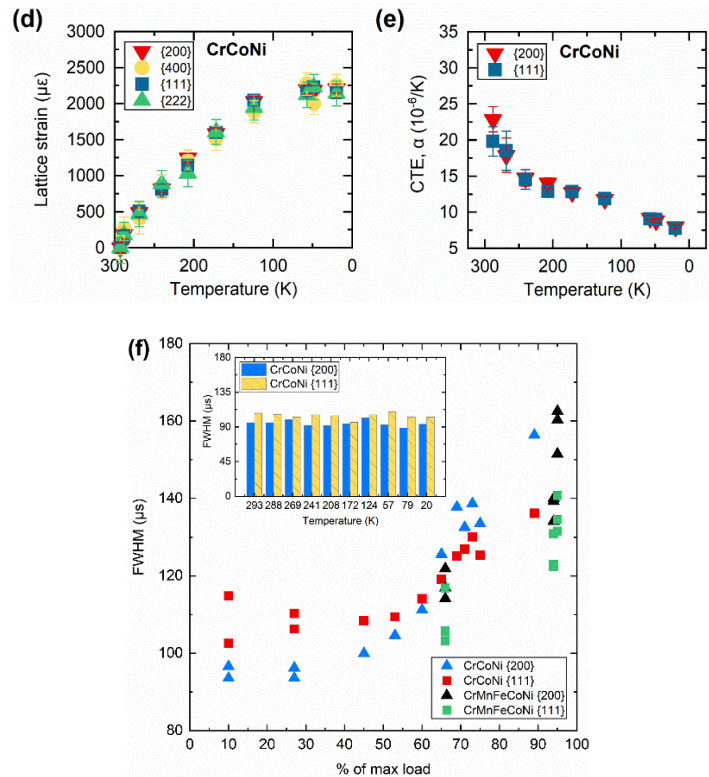

**Supplementary Fig. S5. (a)** Typical examples of ToF neutron diffraction spectra acquired for CrCoNi. Bragg diffraction peaks corresponding to different crystallographic planes have been labeled with *(hkl)*. The magnified view of the lower *d*-spacing planes (in the range of 0.55 to 1 Å) in (a) is plotted in (b) resolved by the dedicated ENGIN-X diffractometer. (c) The measured single phase *fcc* lattice constant $a_0$ in CrCoNi material is 3.562 Å at room temperature and compared to various literature values obtained by experiments or simulations. (d) The calculated lattice strain of {200}, {400}, {111} and {222} planes due to thermal cooling are plotted as a function of temperature. (e) Calculated polycrystalline coefficient of thermal expansion (CTE) as a function of the temperature decrease. (f) Pseudo Voigt profile fitted Full Width at Half Maximum (FWHM) of the diffraction peaks {200} and {111} planes in CrCoNi and CrMnFeCoNi alloys plotted as a function of applied load, the insert in (f) plots CrCoNi diffraction peak FWHM with temperature, showing only loading impacts the peak broadening.

**Supplementary References**